 \documentclass[twocolumn,prb]{revtex4}

\usepackage{graphicx}    % Include figure files
\usepackage{dcolumn}     % Align table columns on decimal point
\usepackage{bm}          % bold math 

\begin{document}

\title{Endohedral terthiophene in zigzag carbon nanotubes: 
Density functional calculations}

\author{W. Orellana$^1$ and Sergio O. V\'asquez$^2$}
\affiliation{$^1$Departamento de F\'{\i}sica, Facultad de Ciencias, 
Universidad de Chile, Casilla 653, Santiago, Chile.\\
$^2$Departamento de Ciencias de los Materiales, Facultad de Ciencias 
F\'{\i}sicas y Matem\'aticas, Universidad de Chile, Av. Tupper 2069, 
Santiago, Chile.}

%date\today

\begin{abstract} 
The inclusion and encapsulation of terthiophene (T3) molecules inside zigzag
single-walled carbon nanotubes (CNTs) is addressed by density functional 
calculations.  We consider the T3 molecule inside six semiconducting CNTs 
with diameters ranging approximately from 8 to 13~\AA. Our results show that 
the T3 inclusion process is exothermic for CNTs with diameters larger than 
9~\AA. The highest energy gain is found to be of 2~eV, decreasing as the CNT 
diameter increases. This notable effect of stabilization is attributed to the 
positively-charged CNT inner space, as induced by its curvature, which is able 
to accommodate the neutral T3 molecule. The band structure of the T3@CNT system 
shows that T3 preserves its electronic identity inside the CNTs, superimposing 
their molecular orbitals onto the empty CNT band structure without hybridization.
Our results predict that the electronic states added by the T3 molecules would 
give rise to optical effects and nonradiative relaxation from excited states.
\end{abstract} 

%pacs{73.22.-f, 71.20.Tx, 31.70.-f}

\maketitle

\section{Introduction}
Filling carbon nanotubes (CNTs) with chosen molecules which in the gas 
phase exhibits particular electronic properties are becoming increasingly 
important owing to the possibility they may modulate CNT properties. Theoretical
investigations have shown that semiconducting CNTs can be amphoterically 
doped in a controllable way by encapsulating electrophilic and nucleophilic 
organometallic molecules \cite{lu}. Other calculations have suggested that
confinement of molecules inside CNTs can induce structural phase transitions 
not seen in the bulk phase. For instance, encapsulated water in an
8~\AA-diameter CNT would exhibit a tight hydrogen-bonding network \cite{hummer} 
as well as acetylene in a 7~\AA-diameter CNT would suffer 
polymerization.\cite{kim} In addition, recent experiments have demonstrated 
that template-grown large-diameter CNTs filled with fluorescent nanoparticles
do emit fluorescent light, allowing the measurement of filling rates and
transport of such nanoparticles inside CNTs.\cite{kim2}

Endohedral functionalization with molecules needs a procedure to open the
nanotubes at a first stage and then fill it with a particular molecule.
Chemical methods are widely used, and the most simple one is using nitric 
acid which is very selective in attacking the carbon pentagon at the tube 
ends.\cite{tsang} Recently, successful methods to fill opened multi-walled
CNTs with fullerenes \cite{frohlich} and others organic molecules \cite{takenobu}
have allowed endohedral inclusion in a soft low-temperature way.
Organic guest molecules like oligo and polythiophenes are an interesting family 
of functional electron systems which have been extensively studied recently due 
to their use in optical devices.\cite{liu,zavelani,bongiovanni} In particular, 
oligothiophenes (T$n$) have spectral properties dependent on the number of 
thiophene units ($n$). They show spectral changes and efficient energy transfer 
processes when included as guests in one-dimensional supramolecular 
architectures, as in perhydrotriphenylene crystals (PHTP).\cite{maniero} This 
characteristic can be exploited for energy conversion or in detectors of 
short-wavelength radiation. Recently, we have reported conformational details 
of the PHTP:T3 inclusion compound.\cite{vasquez}  

In this work, we study the energetic and electronic properties of single-walled 
CNTs filled with terthiophene (T3) molecules as well as the T3 incorporation 
mechanisms by a open-ended CNT.
Single-walled CNTs, which are typically 10~\AA\ in diameter, can be metallic or 
semiconducting depending on the way that a graphene sheet is rolled up, which is 
characterized by the chiral indices ($n,m$). 
Metallic tubes occur if $n-m$ is divisible by 3; otherwise, the tubes are 
semiconducting.\cite{dressel} Non-chiral nanotubes with indices ($n,0$) and ($n,n$) 
are termed $zigzag$ and $armchair$, respectively, which is related to the 
arrangement of the C atoms around the tube. We study the T3 molecule inside six
zigzag CNTs, with diameters larger than 8~\AA. Our results show 
that the inclusion of the T3 molecule is an exothermic process for all CNTs 
under consideration and that the energies involved depend on the diameter of the
nanotube. We also found that the highest occupied and lowest unoccupied molecular
orbitals (HOMO and LUMO) of the free T3 molecule appear as sharp subbands when 
these molecules fill the tube, showing additional singularities in the density of 
states with similar energy difference than that of the free molecule. This result 
suggests that the T3 optical properties could be preserved when this molecule 
fills a semiconducting CNT.

\section{Theoretical method}
The calculations were carried out in the framework of density 
functional theory (DFT),\cite{kohn} with the generalized gradient 
approximation (GGA) to the exchange-correlation functional.\cite{perdew} 
We use a basis set consisting of strictly localized numerical pseudoatomic 
orbitals, as implemented in the SIESTA code,\cite{siesta} namely
double-$\zeta$ singly polarized basis set.
Norm-conserving pseudopotentials \cite{troullier} in their separable 
form \cite{klein} are used to describe the electron-ion interaction.
We consider the zigzag ($n$,0) CNTs with $n$=10, 12, 13, 14, 15, and 16, 
which have diameters of 8.0, 9.6, 10.2, 11.1, 11.4, and 12.7~\AA, respectively. 
Infinite CNTs are described within the supercell approach with
periodic boundary conditions applied only along the nanotube axis.
As the T3 molecule has a length of about 12.4~\AA, we choose a
periodic CNT unit of 17.3~\AA\ in length. Thus, the distance between 
T3 molecules inside a nanotube (T3@CNT) is about 5~\AA, ensuring a 
negligible interaction between them. We consider a lateral distance 
between tubes of about 8~\AA.
We also study the T3 incorporation through an open-ended CNT. This was 
done by taking a finite CNT segment of 17.3~\AA\ plus the T3 molecule 
which is located initially 2~\AA\ apart from the tube end, along its 
axis. Then, the molecule is gradually moved into the tube. We consider 
two possible scenarios for the CNT entrance, with bare C atoms and those
saturated by H atoms. The other CNT end is always saturated by 
hydrogen. 
The positions of all atoms in the nanotubes and the molecule were relaxed 
using the conjugated gradient algorithm until the force components 
become smaller than 0.05~eV/$\text{\AA}$.
Due to the large size of the supercells, we use the $\Gamma$ point for 
the Brillouin zone (BZ) sampling. We check the band-structure calculations 
and the equilibrium geometries for the T3@(12,0) and T3@(14,0) systems 
considering four $k$ points, according to the Monkhorst-Pack mesh.\cite{monk}
Our results do not show any significant variation with respect to the 
$\Gamma$-point calculations. 
We also check the difference in energy between T3@(12,0) and T3@(14,0), 
considering one and four points. We find a very small energy difference
of 3 meV/atom, showing that the $\Gamma$ point is sufficient to ensure 
the convergence of BZ sampling. 

\section{Results and discussion}

%%%%%%%%%%%%%%%%%%%%%%%%%%%%%%%%%%%%%%%%%%%%%%%%%%%%%%%%%%%%%%%%%%
\begin{figure}[t]
\includegraphics[width=6.0cm]{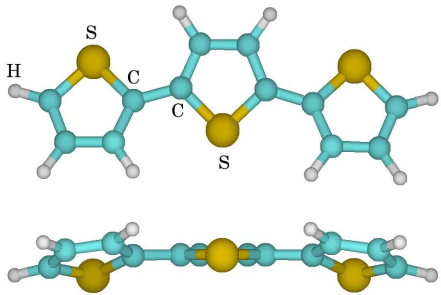}
\caption{(Color on line) Atomic geometry of the T3 molecule in the
gas phase. The three thiophene units adopt a nonplanar configuration
(lower figure) characterized by the dihedral angle S-C-C-S 
($\theta_{\rm D}$).}
\label{f11}
\end{figure}
%%%%%%%%%%%%%%%%%%%%%%%%%%%%%%%%%%%%%%%%%%%%%%%%%%%%%%%%%%%%%%%%%%
We first discuss the structural and electronic properties of the isolated
molecule and the semiconducting CNTs. Gas phase T3 (C$_{12}$H$_{6}$S$_{3}$) has 
been theoretically characterized by a S-C-C-S dihedral angle of 147.2$^{\circ}$,
\cite{dicesare} and experimentally by the $S_0 \rightarrow S_1$ singlet 
transition of 3.05~eV.\cite{yang} Figure~\ref{f11} shows our results for 
the equilibrium geometry; we have obtained a more planar structure 
with a dihedral angle of 157.8$^{\circ}$, whereas the T3 HOMO-LUMO energy is 
found to be 2.33~eV. This discrepancy is a consequence of the DFT-GGA 
methodology which underestimates the gap energy. Concerning the carbon 
nanotubes, we find that (12,0) and (15,0) CNTs are small-gap semiconductors 
with a gap energy of 0.04 and 0.05~eV, respectively, in close agreement with 
experiments,\cite{ouyang} whereas (10,0), (13,0), (14,0) and (16,0) CNTs have 
a larger band gap of 0.85, 0.70, 0.60, and 0.60~eV, respectively. 
%%%%%%%%%%%%%%%%%%%%%%%%%%%%%%%%%%%%%%%%%%%%%%%%%%%%%%%%%%%%%%%%%%
\begin{figure}[t]
\includegraphics[width=8.5cm]{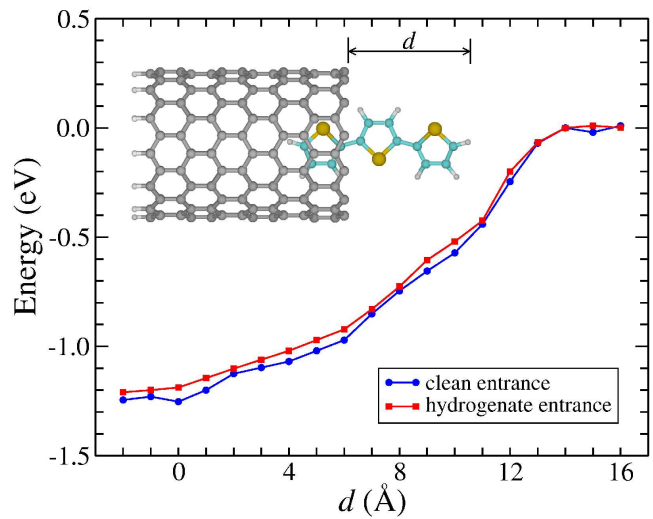}
\caption{(Color on line) Energy profile during the axial insertion 
of the T3 molecule into the (13,0) CNT by both clean and hydrogenate 
entrances. The inset shows the T3 molecule entering inside the 
(13,0) CNT by the clean entrance.}
\label{f22}
\end{figure}
%%%%%%%%%%%%%%%%%%%%%%%%%%%%%%%%%%%%%%%%%%%%%%%%%%%%%%%%%%%%%%%%%%

Figure~\ref{f22} shows the variation in the total energy when T3 enters into a 
(13,0) CNT by a clean and hydrogenated entrance. We find similar energy profiles
in both cases indicating that electrostatic forces induced by the dangling
bonds at the CNT termination are not determinant for the insertion process. 
We observe that the incorporation reaction is exothermic by 1.2~eV. Thus, 
it is energetically favorable that the molecule enters into an uncapped 
nanotube than remains outside, exhibiting a barrierless insertion process.
From the slope of the curves of Fig.~\ref{f22}, we can estimate the capillary 
forces involved in the molecule insertion, considering that $F=\Delta E/\Delta d$. 
We find that the force which pushes T3 into the tube is about 0.15 nN. In a recent
work, Kim {\it et al.} studied the insertion of a C$_2$H$_2$ molecule into 
the (5,5) armchair CNT which has a diameter of about 7~\AA. They found 
a force of 0.22~nN, very close to our results even though the C$_2$H$_2$ 
molecule is smaller and has a linear geometry.\cite{kim} This suggests a 
general trend for the molecular insertion process in carbon nanotubes.

\begin{figure}[t]
\includegraphics[width=8.5cm]{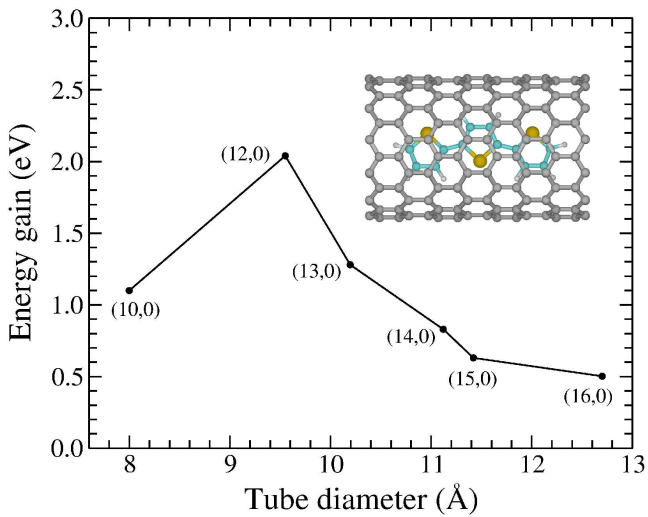}
\caption{(Color on line) Energy gain as a function of the CNT 
diameters for the T3 molecule inside the ($n$,0) CNTs, with $n=10-16$. 
The zero energy corresponds to T3 in the gas phase plus the empty 
CNT. The inset shows the unit cell of T3@(13,0).}
\label{f33}
\end{figure}

We investigate the energetic of zigzag CNTs of different diameters
filled with T3 molecules. In our supercell approach, the molecules 
are aligned coaxially inside the tube with distances separating them 
of about 5~\AA. We find that the energy of the T3@CNT system is lower 
than the isolated constituents. Figure~\ref{f33} shows the gain in
energy of the T3@CNT systems with respect to the empty CNT plus the
isolated T3 molecule. 
%%%%%%%%%%%%%%%%%%%%%%%%%%%%%%%%%%%%%%%%%%%%%%%%%%%%%%%%%%%%%%%%%%
%%%%%%%%%%%%%%%%%%%%%%%%%%%%%%%%%%%%%%%%%%%%%%%%%%%%%%%%%%%%%%%%%%
We find that T3@(12,0) has the highest energy 
gain of about 2~eV, whereas increasing the CNT diameters, this energy
tends to decrease. This notable effect of stabilization is attributed
to the difference in the relative charge between the inner and outer
surfaces of the CNT. For small CNTs, the higher curvature produces
a polarization normal to the tube surface which tend to dislocate the
electronic charge toward outside, as shown by Dumitric\u{a} {et al.},
based on a rehybridization model and DFT calculations.\cite{dumitrica}
Thus, a neutral molecule with a negative electronic cloud surrounding
the positive nucleus would prefer to accommodate in the slightly positive
CNT inner region than remain outside. Therefore, nanocapillarity forces
acting on the T3 molecules would originate in the normal polarization 
between the inner and outer CNT surfaces. For larger-diameter CNTs the 
polarization tends to disappear because the inner and outer curvatures tend 
to equalize. It is worthy to note that the gain in energy for the neutral 
T3 molecule inside CNTs can be as high as 1.5~eV for a small diameter 
variation of about 3~\AA, which is the difference in diameters between 
(12,0) and (16,0) CNTs.

\begin{figure}[t]
\includegraphics[width=8.5cm]{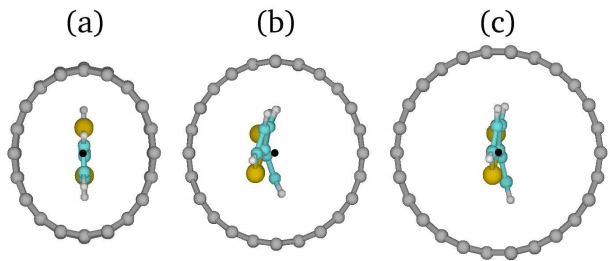}
\caption{(Color on line) Atomic geometry of the T3 molecule inside
carbon nanotubes of different diameters. (a) T3@(10,0), (b) T3@(12,0), 
and (c) T3@(13,0). The black dot indicates the nanotube centers.}
\label{f44}
\end{figure}

Our results for the T3 molecule inside the small (10,0) CNT of 8~\AA\ in 
diameter show a gain in energy of 1.1~eV with respect to the isolated 
components. However, the CNT shows a strong oval deformation whereas the 
molecule becomes completely flat, resulting in a tight encapsulation, as 
shown in Fig.~\ref{f44}(a). Although the (10,0) CNT has a stronger normal 
polarization owing to its enhancing curvature, the T3 inclusion process 
appears to be unrealistic due to steric effects. The above result 
suggests that benzenelike molecules would not enter into uncapped zigzag 
CNTs with diameters closer to 8~\AA.

Table~\ref{t1} compares the geometry and electronic properties of T3 in 
the gas phase and inside the CNTs. We find a larger molecular deformation 
in T3@(12,0), reducing the diahedral angle in $14^{\circ}$ with respect 
to the free molecules. 
This tendency reverses when the CNT diameter increases. Figures~\ref{f44}(b) 
and \ref{f44}(c) show an axial view of the T3@(12,0) and T3@(13,0) systems. 
We observe that the T3 molecule inside the (12,0) CNT is shifted about 
0.5~\AA\ toward the wall from the CNT axis with the S atoms facing the wall, 
whereas inside the (13,0) CNT the molecule remains at the CNT axis, 
increasing the dihedral angle. This result somewhat confirms the 
positive character of the tube inner wall which in the (12,0) CNT induces 
a sufficiently high normal polarization that moves the molecule toward the 
wall, decreasing the dihedral angle. Table~\ref{t1} also shows that 
both the CNT band gap and the T3 HOMO-LUMO energy remain essentially 
unchanged when the molecule is inside the CNTs. 
%%%%%%%%%%%%%%%%%%%%%%%%%%%%%%%%%
 \begin{table}[b]
 \caption{\label{t1} Structural and electronic parameters for the T3 
  molecule and the T3@CNT system. $\theta_{\rm D}$ is the dihedral 
  angle, and $E_{\rm HL}$ is the HOMO-LUMO energy of the molecule.
  $E_{\rm G}$ is the T3@CNT gap energy.}
 \begin{ruledtabular}
 \begin{tabular}{ccccc}    
  System & $\theta_{\rm D}$ (deg) & $E_{\rm HL}$ (eV) & $E_{\rm G}$ (eV) \\
 \hline
   T3        & 157.8 &  2.33 & -    \\
   T3@(12,0) & 143.8 &  2.30 & 0.05 \\
   T3@(13,0) & 157.3 &  2.29 & 0.70 \\
   T3@(14,0) & 155.0 &  2.34 & 0.60 \\
   T3@(15,0) & 167.8 &  2.34 & 0.05 \\
   T3@(16,0) & 161.8 &  2.27 & 0.60 \\
 \end{tabular}
 \end{ruledtabular}
 \end{table}
%%%%%%%%%%%%%%%%%%%%%%%%%%%%%%%%%  

%%%%%%%%%%%%%%%%%%%%%%%%%%%%%%%%%%%%%%%%%%%%%%%%%%%%%%%%%%%%%%%%%%
%%%%%%%%%%%%%%%%%%%%%%%%%%%%%%%%%%%%%%%%%%%%%%%%%%%%%%%%%%%%%%%%%%

\begin{figure}
\includegraphics[width=8.0cm]{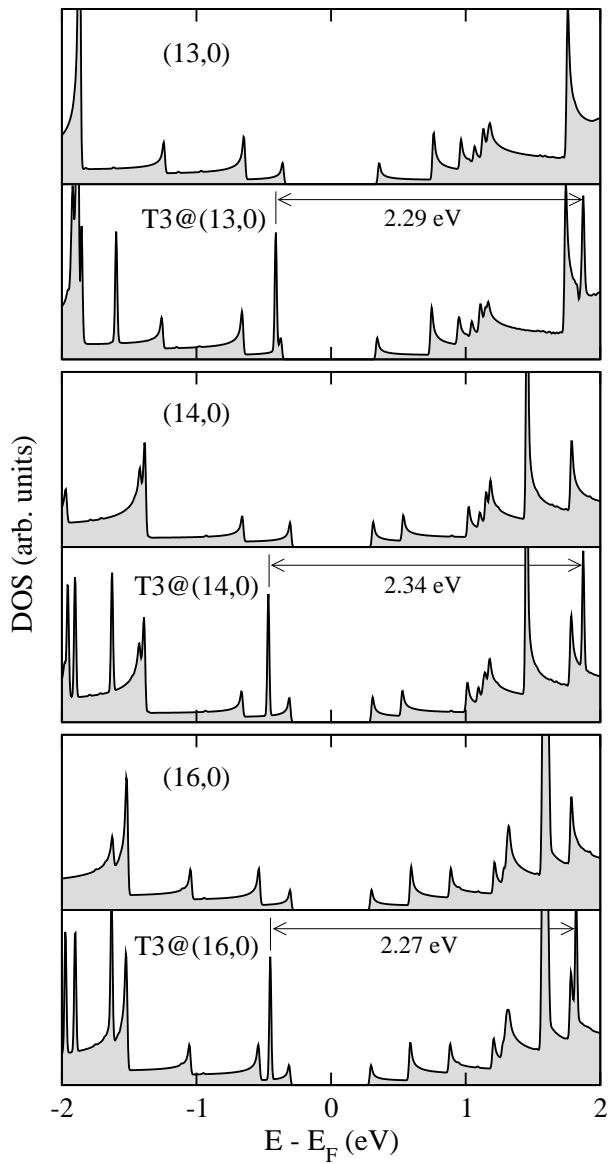}
\caption{Density of states (DOS) of empty CNTs (13,0), (14,0),
and (16,0) (upper panels) and DOS of the same CNTs filled with T3 
molecules (lower panel). The arrows indicate the HOMO-LUMO energy
width of T3.}
\label{f55}
\end{figure}

Figure~\ref{f55} shows the density of states (DOS) of the empty CNTs (13,0), (14,0), 
and (16,0) (upper panels), which are compared with the DOS of the same CNTs 
filled with T3 molecules (lower panels). The DOS of filled CNTs clearly 
shows the appearance of both HOMO and LUMO states of T3 at essentially the 
same energy, about $E_F - 0.44$~eV and $E_F + 1.86$~eV, respectively. 
The difference in energy between these states is found to be of 
2.30~eV, very close to that calculated for the free molecule, 2.33~eV. Other molecular
states can be observed between $E_F - 2.0$~eV and $E_F - 1.5$~eV.
This suggests a minor influence of the inner CNT electrostatic potential in 
the T3 molecular states. Therefore, we may speculate that the optical properties 
of the T3 molecule would be maintained inside the tube. 
Nevertheless, the addition of molecular states to the band structure of 
the CNT itself produces a system with more complex optical characteristics,
where we may predict fluorescence from the T3 molecule as well as fluorescence 
due to transitions between the T3 HOMO band to the sharpest Van Hove singularity 
in the conduction band.
%%%%%%%%%%%%%%%%%%%%%%%%%%%%%%%%%%%%%%%%%%%%%%%%%%%%%%%%%%%%%%%%%%
%%%%%%%%%%%%%%%%%%%%%%%%%%%%%%%%%%%%%%%%%%%%%%%%%%%%%%%%%%%%%%%%%%
In addition, highly probable energy transfer processes between oligothiophene 
molecules and the CNT may occur. In this case, a plausible mechanism would 
allow absorption in the near UV ($S_0 \rightarrow S_1$ of T3), with energy transfer 
from T3 (acting as donor) to the CNT (acting as acceptor). For instance, 
down-conversion processes between Van Hove singularities would give fluorescent 
emission from $E_{33}$ up to the $E_{11}$ IR transition, in the case of the T3@(13,0) 
system. It is important to remember that the calculated energy transitions were 
obtained within the DFT-GGA approach which underestimates their values. 
Thus, the optical properties extracted from the above results must be taken as a
qualitative picture.

\section{Summary and conclusions}

A notable effect of stabilization is produced by the endohedral inclusion
of the oligomer through open-ended zigzag single-walled CNTs, which is the 
order of magnitude of a chemical bond. This effect is observed with similar 
magnitude in both bare and hydrogenated CNT entrances. Earlier studies of 
several neutral fullerenes predict that the electrostatic potential inside 
the cage has a positive charge character and consequently would be suitable 
to accommodate anionic or neutral species.\cite{cioslowski} More recent 
calculations confirms this finding in carbon nanotubes,\cite{dumitrica} 
showing that the curvature of the CNT walls produces a charge displacement 
toward the outside, leaving a positive character inside. Thus, the T3 inclusion 
or encapsulation in a CNT is driven in a significant extent by nanocapillarity 
forces induced by the interaction between the slightly positive CNT cavity 
and the electronic cloud of the molecule, giving an exothermic process. 
Thus, the drop in the energy gain for the T3 inclusion in larger-diameter 
CNTs can be explained in terms of their decreasing curvature.

The band structures of T3@CNTs show that the T3 molecule preserves its
electronic properties, superimposing the HOMO-LUMO states onto the original 
CNT band structure. Semiconducting single-walled CNTs are optically active
systems that are fluorescent when encased in cylindrical micelles with a 
striking influence on the diameter and chiral angle,\cite{bachilo,kono}
showing strong absorption and fluorescence due to their sharp Van Hove
singularities. Our results predict that the electronic states added by 
the T3 molecules to the CNT band structure would give rise to optical effects
in the radiative relaxation from the excited states. In addition, energy-transfer 
processes between T3 and the CNT suggest possible applications
in optical nanodevices, for instance, in UV down-converters.

\acknowledgments
This work was supported by the Chilean founding agencies
FONDECYT under grant Nos. 1050197 (W.O.) and 1030662 (S.O.V.) 
and the Millennium Nucleus of Applied Quantum Mechanics and 
Computational Chemistry through project No. P02-004-F.

\end{document}